\documentclass{PoS}
\usepackage{appendix}

\usepackage{pdflscape}
\usepackage{afterpage}

\usepackage[symbol]{footmisc}

\title{Searching for Time-Dependent Neutrino Emission from Blazars with IceCube}

\ShortTitle{Searching for Time-Dependent Neutrino Emission from Blazars with IceCube}

\author{
The IceCube Collaboration\footnote{For collaboration list, see PoS(ICRC2019) 1177.}\\
{\itshape \href{http://icecube.wisc.edu/collaboration/authors/icrc19_icecube}{http://icecube.wisc.edu/collaboration/authors/icrc19\_icecube}}\\
E-mail: \email{erin.osullivan@fysik.su.se, cfinley@fysik.su.se}
}

\abstract{

In 2017, IceCube detected a high energy neutrino in coincidence with the blazar TXS 0506+056. In a follow up analysis of archival data, evidence for previous neutrino emission from the blazar during 2014-15 was found. In this analysis, we investigate whether other blazars might have had episodes of neutrino emission. We use a 5-year sample of muon-neutrino data and the same time-dependent algorithm that was used for the archival analysis of TXS 0506+056, applying it to all of the objects in the northern sky listed in the Fermi 3LAC catalog. We fit for the most significant neutrino flare from each direction, without assuming any correlation with electromagnetic light curves. To search for an excess of significant flares which could indicate a small population of neutrino emitters, we apply a binomial test. This binomial test does not find a significant excess of flares relative to the results expected for background coincidences from atmospheric neutrinos. The final significance of the analysis is p=0.244 when TXS 0506+056 is excluded, consistent with the background-only hypothesis. We report the parameters of the individual fits for the most significant objects which contributed to the final result. 

\vspace{4mm}
{\bfseries Corresponding authors:}
\speaker{Erin O'Sullivan}$^{1}$, Chad Finley$^{1}$\\
{$^{1}$ \itshape Oskar Klein Centre and Dept. of Physics, Stockholm University}\\

}

\FullConference{36th International Cosmic Ray Conference -ICRC2019-\\
		July 24th - August 1st, 2019\\
		Madison, WI, U.S.A.}

\begin{document}

\section{Introduction}\label{sec:intro}

On September 22, 2017, IceCube detected a high energy neutrino from the direction of a flaring blazar TXS 0506+056 seen by the Fermi LAT \cite{IceCube_blazar_Science18}. Analysis of the archival data from the direction of TXS 0506+056 revealed a 3.5$\sigma$ flare of neutrinos between September 2014 and March 2015 that appears uncorrelated with enhanced gamma emission \cite{IceCube_archival_blazar_Science18}. Taken together, these two findings point to blazar TXS 0506+056 being a likely source of high energy neutrinos. 

IceCube \cite{Aartsen16} is a cubic-kilometer-scale neutrino detector embedded in the ice at the South Pole. Because of the small cross section for neutrino interactions, a large
detector volume is required to detect astrophysical neutrinos. The IceCube detector consists of 5160 optical sensors positioned over 86 strings, at depths ranging from 1.5-2.5 km below the surface of the ice. A summary of recent results from the experiment are described in \cite{Williams:2019icrc_IC}. 

The analysis described in this proceeding was designed to further understand the connection between neutrinos and blazars by searching for time-dependent neutrino emission detected in IceCube from the directions of blazars and other AGNs listed in the Fermi 3LAC catalog \cite{3LAC15}. Here, we report the individual fit results for the most significant objects that contributed to the final result and perform a binomial test to look for statistically significant flares from a small number of sources in the catalog.

\section{Analysis method}

\begin{figure}[h!]
\centering
\includegraphics[trim={1cm 2cm 1cm 1cm},clip,width=.85\textwidth]{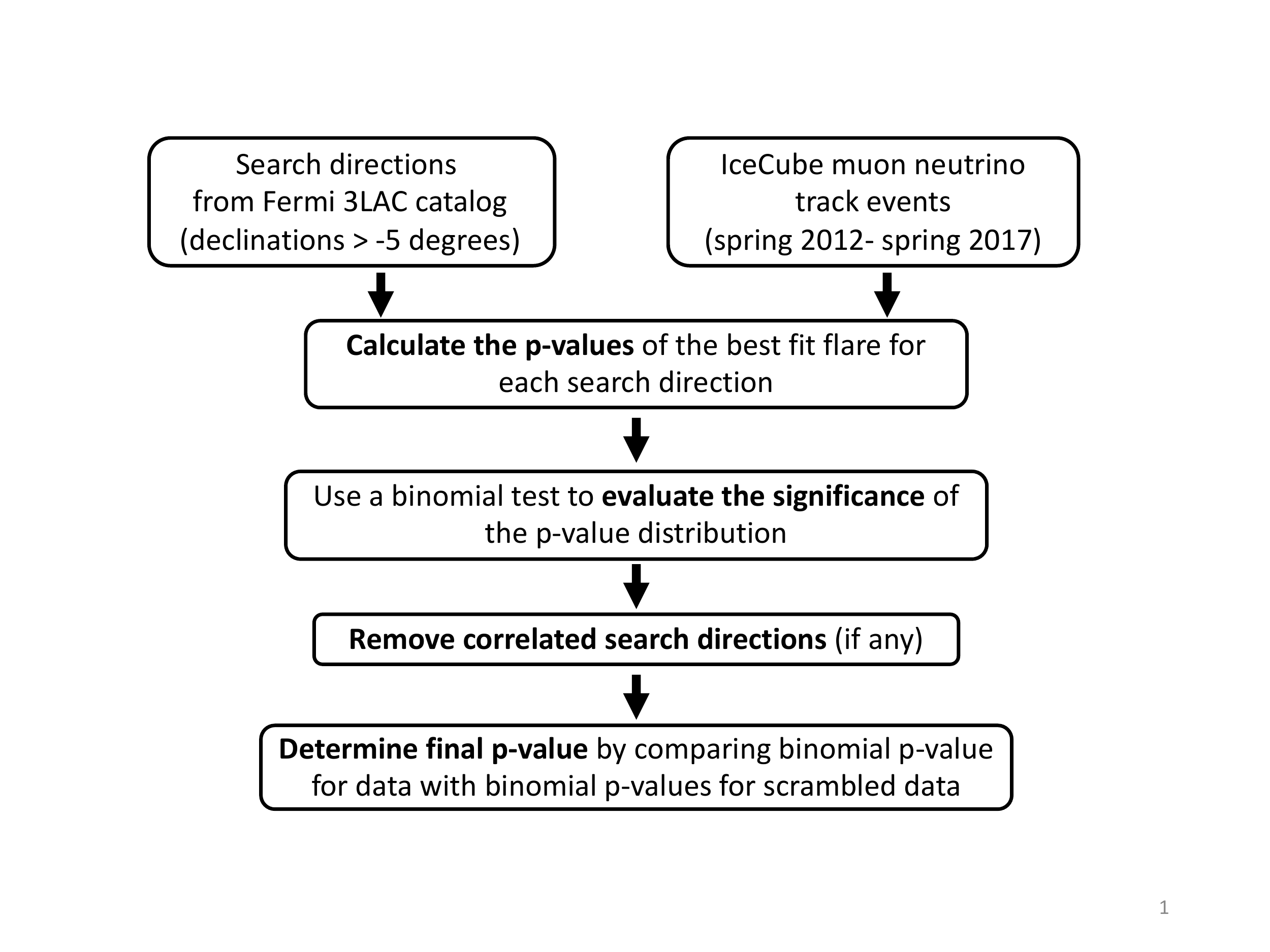}
\caption{A flowchart showing the steps of this analysis.}
\label{fig1}
\end{figure}

The outline of this analysis is shown in Figure \ref{fig1}. We used the data set described in \cite{Aartsen18diffuse},  consisting of well-reconstructed muon tracks from atmospheric and astrophysical neutrinos in the time period from April 26, 2012 to May 11, 2017. We used the direction of the 1023 objects in the Fermi 3LAC catalog \cite{3LAC15} with declinations above -5$^\circ$, and report the details of the most significant neutrino flare for each direction using the unbinned maximum likelihood method described in Section \ref{sec:code}. We then perform the binomial test, as described in Section \ref{binomial_test} on the set of p-values to assess the statistical significance of the ensemble. In order to account for the fact that the results for nearby blazar directions can be influenced by the same neutrino events, we remove correlated results using a prescription described in Section \ref{correlations}. Finally, we compare the decorrelated p-value from the binomial test with that obtained from scrambled data to calculate a final p-value. 

\subsection{Unbinned maximum likelihood method}\label{sec:code}

To find the most significant neutrino flare from each search direction, we used the unbinned maximum likelihood method described in \cite{Braun08} and \cite{Braun09}. 

The best fit flare is found by maximizing the likelihood function, 
\begin{equation}
    L = \prod_{i=1}^{N} \left[\frac{n_s}{N}S_i+(1-\frac{n_s}{N})B_i\right]
\end{equation}
where $N$ is the number of neutrino events in the sample, $n_s$ is the number of fitted signal events that maximizes the likelihood, and $S_i$ and $B_i$ are the signal and background PDFs. The signal and background PDFs are constructed using spacial, energy, and temporal information.

The signal PDF can be written as
\begin{equation} \label{S_i}
    S_i=\frac{1}{2\pi\sigma_i^2}e^{-\frac{r_i^2}{2\sigma_i^2}} \times P(E_i|\gamma) 
    \times \frac{1}{\sqrt{2\pi} T_{\sigma}}e^{-\frac{(T_i - T_O)^2}{2T_{\sigma}^2}}.
\end{equation}
The first term represents the spatial component of the likelihood, where reconstructed neutrino event $i$ has an angular resolution $\sigma_i$ and an angular separation with the search direction $r_i$. The second term, $P(E_i|\gamma)$, gives the probability distribution for the reconstructed energy values E$_i$ given a power-law energy spectrum with index $\gamma$. Finally, the last term accounts for the timing information. We weight the neutrino events according to a Gaussian profile, where the central time of the peak, as well as the width of the peak, are fit. Here, T$_{\sigma}$ is the half width of the time window, $T_i$ is the time of the neutrino event, and $T_o$ is the centre of the time window.

The background PDF can be similarly constructed, where
\begin{equation} \label{B_i}
    B_i = \frac{1}{\Omega T_L}\times P_{atm}(E_i) 
\end{equation}
with $\Omega$ representing the solid angle of the declination band of the search direction containing $N$ total events, $T_L$ representing the total livetime of the search, and $P_{atm}(E_i)$ representing the probability of getting an event with energy $E_i$ given the atmospheric neutrino energy spectrum. 

Our final test statistic, which includes a leading term to penalize short time windows as they should have a larger trial factor, is
\begin{equation}
    D = -2 \log\left[\frac{T_{L}}{\hat{T}_\sigma} \times \frac{L(\hat{n}_s = 0)}{L(\hat{n}_s, \hat{\gamma}, \hat{T}_o, \hat{T}_\sigma)}\right]
\end{equation}
where $\hat{n}_s, \hat{\gamma}, \hat{T}_o,$ and $ \hat{T}_\sigma$ are the parameters that maximize the likelihood. 

This test statistic is converted to a p-value for the search direction by comparing the value obtained from data with the distribution of the test statistic from scrambled data with randomized positions and times.  

\subsection{Binomial test}\label{binomial_test}

Once we have obtained a p-value for the best fit for a potential neutrino flare in each of the search directions, we use the binomial test to determine if the p-values are lower than expected from a random distribution of such p-values. The binomial test was chosen as it is optimized to search for a small number of neutrino emitters within a larger population. 

In order to use the binomial test to determine the significance of the population of p-values, we order the p-values from lowest to highest and scan over the 100 lowest p-values from $k$=1 (the lowest p-value) to $k$=100 (the 100th lowest p-value). The probability of getting $k$ or more blazars at or below local p-value $p_k$, given the total number $N$ of p-values in the set is
\begin{equation}
    P(k) = \sum_{m=k}^N \frac{N!}{(N-m)!m!}p_k^m(1-p_k)^{(N-m)} 
\end{equation}
After scanning over the top 100 most significant directions, we report the lowest binomial p-value and the number of blazars it corresponds to, $k$. 

\subsection{Accounting for correlated p-values in the binomial test}\label{correlations}

The binomial test assumes that the p-values are distributed uniformly between 0 and 1. In our case, the binomial test has artifacts due to correlations, 
making the binomial p-value, $P(k)$ not uniformly distributed even for 
scrambled data. These correlations arise from nearby blazar directions that are influenced by the same nearby neutrino events. 

The impact of potentially correlated p-values was addressed with the 
following method: We define two blazar directions to be correlated if the same neutrino significantly contributes to both of the neutrino flares. We define the significance of neutrino $i$ as the ratio of $\frac{S_i}{B_i}$ where $S_i$ is defined in equation \ref{S_i} and $B_i$ is defined in equation \ref{B_i}. We rank the neutrinos using this weighting and take the top $\hat{n}_s \times 3$ neutrinos as being significant contributors to the fit. If two blazar directions share a significantly contributing neutrino, they are defined as correlated directions. 

To remove the correlations, we determine how many AGNs in the top $k$ need to be removed such that only uncorrelated directions remain in this list. We then reduce $k$ by the number of AGNs we need to remove, for example if we find a set of three correlated directions, we would reduce $k$ by two. We then recalculate the binomial p-value, keeping the local p-value at the original best fit $k$. In this way, there is no need to decide {\it which} blazar or blazars to remove, we simply ensure that the final significance is not artificially enhanced by double counting.

We determine the final p-value by repeating this prescription on scrambled data sets and comparing the frequency of our decorrelated binomial p-value with the p-values obtained from these scrambles. 

\section{Results}

\subsection{Full catalog}

\begin{figure}[h!]
\centering
\includegraphics[trim={2cm 1cm 1cm 2cm},clip, width=.65\textwidth, angle=-90]{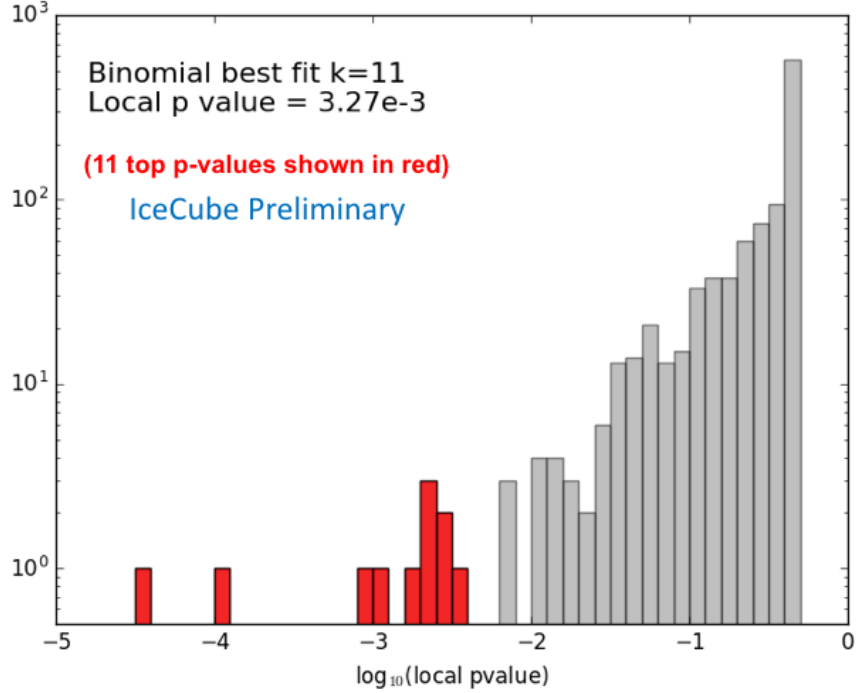}
\caption{The distribution of p-values for the most significant neutrino flare in the direction of all 1023 northern-sky objects in the 3LAC catalog. The 11 p-values which contribute to the most significant binomial test result are shown in red.}
\label{local_p_vals}
\end{figure}

The most significant neutrino excess was found for each direction in the Fermi 3LAC catalog. Figure \ref{local_p_vals} shows the distribution of p-values found from each direction in the catalog. Performing the binomial test as described in Section \ref{binomial_test}, we found the smallest binomial test p-value at k=11. The p-values for the top 11 3LAC directions are shown in red in Figure \ref{local_p_vals}. A table with the catalog and fit information for the top 11 search directions can be found in Table \ref{tab1}. 

We found two sets of correlations in the top 11 most significant search directions in the data, one involving three search directions (corresponding to counterparts B2 1126+37, MG2 J112910+3702, and MG2 J112758+3620) and one involving two (corresponding to counterparts GB6 J0929+5013 and 1ES 0927+500). In both cases, the correlated search directions were spatially close and the fit was clearly influenced by the same neutrinos. In order to remove correlations from the list, we reduced the k-value by three (two for the correlation involving three search directions and one for the correlation involving two) and recalculated the binomial test using $k=8$, but keeping the local p-value at 3.27$\times$10$^{-3}$. This decorrelated binomial p-value was found to be 2.09$\times$10$^{-2}$. We also performed the calculation with $k=7$ to remove TXS 0506+056 as this result was already known and provided motivation for this analysis. The decorrelated binomial probability without TXS 0506+056 was 5.33$\times$10$^{-2}$. 

\begin{figure}[h!]
\centering
\includegraphics[trim={2.5cm 1cm 1cm 3cm},clip, width=.6\textwidth, angle=-90]{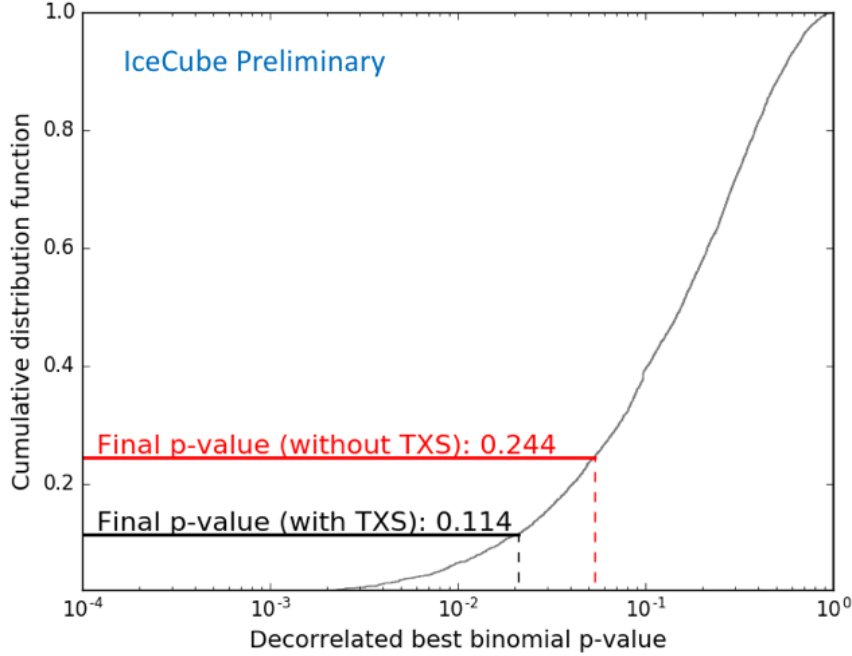}
\caption{The distribution of binomial test p-values, with correlations removed, from analysis of scrambled data (grey), compared with the value obtained in the data, including TXS 0506+056 (black) and without TXS 0506+056 (red).}
\label{fig:scan_correction}
\end{figure}

\begin{figure}[h!]
\centering
\includegraphics[trim={4cm 2cm 4cm 2cm},clip, width=.55\textwidth, angle=-90]{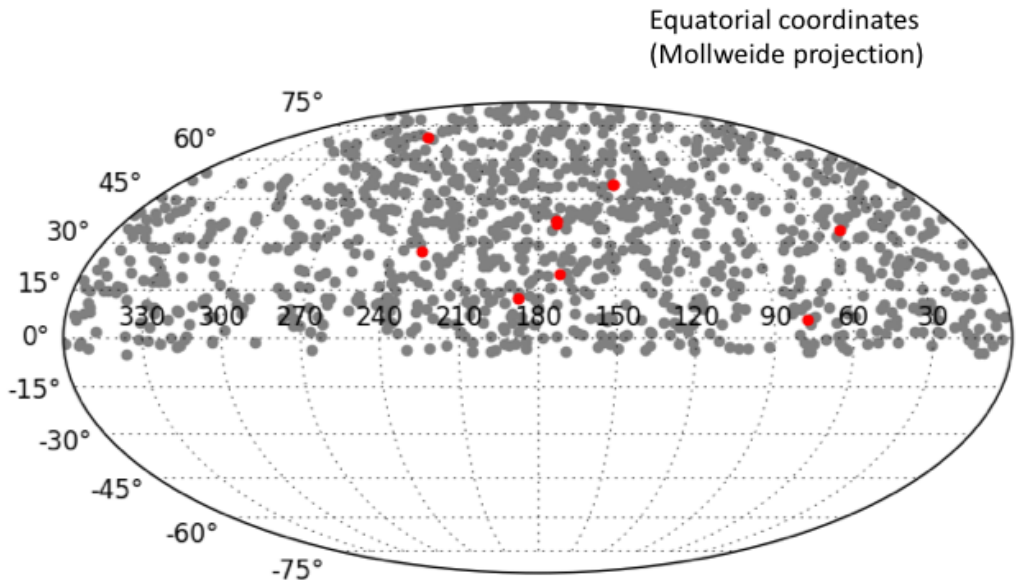}
\caption{Skymap in equatorial coordinates of the search directions in the 3LAC catalog, shown in grey. In red, we show the 11 search directions that yield the lowest p-value in the binomial test (8 of which are unique).}
\label{fig:skymap}
\end{figure}

The final p-value for this analysis is calculated by comparing the decorrelated p-value obtained from data to the distribution of decorrelated p-values obtained from scrambled data. Figure \ref{fig:scan_correction} shows the decorrelated p-value, both with and without TXS 0506+056, compared with the distribution from scrambled data. The final corrected p-value is 0.114 with TXS 0506+056 and 0.244 without. Figure \ref{fig:skymap} shows the positions of the search directions on a skymap.

\subsection{Testing the Optical Class Sub-catalogs}

We separately tested two sub-catalogs of the 3LAC catalog: directions corresponding to blazars classified as FSRQ, and directions corresponding to blazars classified as BLLac. Using the full statistical test described above (performing the binomial test, decorrelating, and comparing to scrambled data), we found for the FSRQ-only sub-catalog that the best k is 45 with a final p-value of 0.50 and for the BLLac-only sub-catalog we found the best k is 42 with a final p-value of 0.60.

It can be noted in Table 1 that 3 directions in our top 11, including 2 of our top 3, have non-blazar counterparts.  Such objects make up a small fraction of the 3LAC catalog, with only 26 non-blazar classifications out of the 1023 counterpart objects in the northern sky. While this should not be considered significant evidence due to the {\it a posteriori} nature of this observation, the result suggests it may be worth further investigating the objects in this sub-catalog.

\section{Conclusions}

Using the same analysis method that found time-dependent neutrino emission from TXS 0506+056, we search in the directions of Fermi 3LAC objects in the northern sky for other blazars with neutrino emission. We report in these proceedings the individual most significant fit results. A binomial test was used to determine whether there were more high-significance neutrino flares than one expects from chance coincidence of background neutrinos. The test did not indicate such an excess for the full catalog nor for the FSRQ and BL Lac sub-catalogs.   

\bibliographystyle{ICRC}
\bibliography{references}

\appendix

\begin{landscape}
\begin{table}[]
    \centering
\begin{tabular}{c c c c c c c c c c}

	 Name  &  Counterpart  &  Optical class &  RA [$^\circ$] &  Dec [$^\circ$] & $\hat{n}_s$ & $\hat{\gamma}$ & $\hat{T}_{o}$ [MJD]& $\hat{T}_W$ [days] & p-value \\ 
	 \hline
	  3FGL J0509.4+0541   &   TXS 0506+056   &   bll   & 77.36 & 5.69 & 12.3 & 2.2 & 57000 & 1.2$\times$ 10$^2$ & 3.47$\times$10$^{-5}$ \\ 
	  3FGL J0325.2+3410   &   1H 0323+342   &   nlsy1   & 51.17 & 34.18 & 2.0 & 1.7 & 57326.2938 & 1.7$\times$10$^{-3}$ & 1.00$\times$10$^{-4}$ \\ 
	  3FGL J1129.0+3705   &   B2 1126+37\footnote[1]{correlated triplet}   &   agn   & 172.29 & 37.15 & 4.0 & 3.3 & 56501.385 & 6.0$\times$10$^{-2}$ & 9.56$\times$10$^{-4}$ \\ 
	  3FGL J1129.0+3705   &   MG2 J112910+3702\footnote[1]{correlated triplet}   &   bll   & 172.31 & 37.05 & 4.0 & 3.3 & 56501.385 & 6.0$\times$10$^{-2}$ & 1.01$\times$10$^{-3}$ \\ 
	  3FGL J1230.9+1224   &   M 87   &   rdg   & 187.71 & 12.39 & 3.0 & 3.4 & 57730.0307 & 2.7$\times$10$^{-3}$ & 1.91$\times$10$^{-3}$ \\ 
	  3FGL J1127.8+3618   &   MG2 J112758+3620\footnote[1]{correlated triplet}   &   fsrq   & 172.00 & 36.34 & 4.0 & 3.3 & 56501.386 & 6.0$\times$10$^{-2}$ & 2.03$\times$10$^{-3}$ \\ 
	  3FGL J0929.4+5013   &   GB6 J0929+5013\footnote[2]{correlated doublet}   &   bll   & 142.31 & 50.23 & 5.3 & 1.9 & 57758.0 & 1.2 & 2.26$\times$10$^{-3}$ \\ 
	  3FGL J1715.7+6837   &   S4 1716+68   &   fsrq   & 259.06 & 68.61 & 2.0 & 4.0 & 57469.17919 & 5.4$\times$10$^{-5}$ & 2.36$\times$10$^{-3}$ \\ 
	  3FGL J1125.9+2007   &   4C +20.25   &   fsrq   & 171.49 & 20.10 & 5.7 & 2.6 & 56464.1 & 5.2 & 2.79$\times$10$^{-3}$ \\ 
	  3FGL J1508.6+2709   &   RBS 1467   &   bll   & 227.18 & 27.15 & 17.3 & 2.9 & 57440 & 1.7$\times$10$^{2}$ & 2.84$\times$10$^{-3}$ \\ 
	  3FGL J0930.0+4951   &   1ES 0927+500\footnote[2]{correlated doublet}   &   bll   & 142.66 & 49.84 & 5.4 & 2.0 & 57758.0 & 1.2 & 3.27$\times$10$^{-3}$ \\ 

\end{tabular}
\caption{The catalog information and best-fit neutrino flare information for the top 11 directions from the 3LAC catalog, 8 of which are unique. In this list, there is one set of three correlated search directions (indicated in the table with an *: B2 1126+37, MG2 J112910+3702, and MG2 J112758+3620) and one set of two (indicated in the table with a $\dagger$, GB6 J0929+5013 and 1ES 0927+500). The optical class descriptions can be found in \cite{3LAC15}. $\hat{T_W}$ represents 2 $\times$ $\hat{T_{\sigma}}$, the half-width of the time window.}
\label{tab1}
\end{table}
\end{landscape}



%

\end{document}